\documentclass[superscriptaddress,groupedaddress,nofootnoteinbib,12pt]{article}  

\usepackage{graphicx}  
\usepackage{dcolumn}   
\usepackage{bm}        
\usepackage{amssymb}   
\usepackage{jcappub}

\def\be{\begin{equation}}
\def\ee{\end{equation}}
\def\ba{\begin{eqnarray}}
\def\ea{\end{eqnarray}}
\def\beq{\begin{eqnarray}}
\def\eeq{\end{eqnarray}}

\def\mpl{M_{\rm Pl}}
\def\e{{\epsilon}}

\def\d{\mathrm{d}}
\def\p{{\cal P}}

\def\K{{\cal K}}

\def\L*{{\cal L}_*}
\def\L{\mathcal{L}}
\def\({\left(}
\def\){\right)}

\def\nn{\nonumber}
\def\p{\partial}
\def\mn{_{\mu \nu}}
\def\stu{St\"uckelberg }
\def\p{\partial}

\def\<{\langle}
\def\>{\rangle}

\def\K{\mathcal{K}}

\def\cs2{c_{s}^{2}}

 \def\al{\alpha}
 \def\b{\beta}
 
 \def\de{\delta}
 
 \def\ep{\varepsilon}

 \def\La{\Lambda}
 \def\si{\sigma}

 \def\om{\omega}

 \def\p{\partial}

 \def\be   {\begin{equation}}   \def\ee   {\end{equation}}

 \def\ba  {\begin{eqnarray}}   \def\ea  {\end{eqnarray}}

\begin{document}
\title{Complete Decoupling Limit of Ghost-free Massive Gravity}

\author{Nicholas A. Ondo, Andrew J. Tolley}
\affiliation{Department of Physics, Case Western Reserve University, 10900 Euclid Ave, Cleveland, OH 44106, USA}
\date{\today}


\abstract{ We present the complete form of the decoupling limit of ghost-free massive gravity with a Minkowski reference metric, including the full interactions of the helicity-1 and helicity-0 modes of the massive spin-2 field.  While in the metric language the square root structure of the mass terms makes it difficult to find a simple way to write down the interactions, we show that using the vierbein formulation of massive gravity, including \stu fields for both diffeomorphism and local Lorentz symmetries, we can find an explicitly resummed expression for the helicity-1 field interactions. We clarify the equations of motion for the Lorentz \stu fields and how these generate the symmetric vierbein condition which guarantees equivalence between the vierbein and metric formulations of massive gravity.
}

\maketitle


\section{Introduction}
\label{sec: intro}

Due to the discovery of cosmic acceleration, and the renewed importance of addressing the cosmological constant problem, there has recently been a revival of interest in infrared modified theories of gravity, such as massive gravity. Historically, a consistent theory was only known to linear level, the so-called Fierz-Pauli theory which was discovered over 70 years ago,~\cite{Fierz:1939ix}. This theory correctly describes a linearized massive spin-two field which at high energies can be decomposed into 2 helicity-2 modes, 2 helicity-1 modes and 1 helicity-0 mode. It was later discovered that there were obstructions to the na\"ive versions of the non-linear, interacting theories of massive gravity, both from the existence of the vDVZ discontinuity \cite{vanDam:1970vg} (i.e. the failure to recover GR in the limit $m \rightarrow 0$) and because of the existence of a spurious degree of freedom, the Boulware-Deser ghost \cite{ Boulware:1973my}. The vDVZ discontinuity was resolved by Vainshtein \cite{Vainshtein:1972sx} who recognized that Einstein gravity is recovered in the massless limit of massive gravity provided only that proper account is taken of the nonlinearities due to the additional degrees of freedom. The mechanism by which the nonlinearities of the helicity-0 mode of the massive graviton can screen the existence of fifth forces has since become known as the Vainshtein mechanism. 
Recently it has emerged that there is a consistent non-linear extension, in which Einstein gravity is supplemented by the a specific choice of nonlinear mass terms (henceforth dRGT mass terms), that evades the Boulware-Deser ghost which famously plagued massive gravity theories \cite{deRham:2010ik, deRham:2010kj, deRham:2011rn, deRham:2011qq}:
\ba
\label{eq:metricaction}
S_{\rm metric}&=& \frac{\mpl^2}{2} \int d^4 x \sqrt{g} \(R +2 m^2 [e_2(\mathcal{K}) + \alpha_3 e_3(\mathcal{K}) + \alpha_4 e_4(\K) ] \)\,,
\ea
where the potential terms $e_{n}(\K)$ are defined as follows:
\ba
&& e_2(\K) = \frac{1}{2!}\( [\K]^2 - [\K^2] \) \nonumber \\
&& e_3(\K) = \frac{1}{3!}\( [\K]^3 - 3[\K][\K^2] + 2[\K^2] \) \nonumber \\
&& e_4(\K) = \frac{1}{4!} \( [\K]^4 - 6[\K]^2[\K^2] + 3[\K^2]^2 + 8[\K][\K^3] -6[\K^4] \)\,,\nn
\ea
with
\ba
\K^{\al}_{\b}= \de^{\al}_{\b} - \( \sqrt{g^{-1}f }\)^{\al}_{\b}\,,
\ea
$f\mn$ being the reference metric.
Initially the theory was shown to be ghost-free both in the decoupling limit \cite{deRham:2010ik,deRham:2010kj} and perturbatively to fourth order --where the decoupling limit is a specific scaling limit of dRGT where the interactions take on a simple high-energy form. Since then, it has exhaustively been shown to be ghost free outside the decoupling limit and to all orders with many methods \cite{deRham:2011qq, Hassan:2011hr, Hassan:2011tf, Hassan:2011ea, Mirbabayi:2011aa, Golovnev:2011aa, Hassan:2012qv, Deffayet:2012nr}.  Subsequently, dRGT has been extended to the vierbein formalism \cite{Hinterbichler:2012cn,Chamseddine1} (for prescient earlier work see \cite{Chamseddine2,Nibbelink:2006sz}), which can more succinctly demonstrate the absence of the BD ghost and greatly simplifies the form of the interactions.

Given that the full form of ghost-free massive gravity is now known, one may wonder why it is necessary to further consider the decoupling limit theory. In practice the majority of the phenomenological understanding about massive gravity and the understanding of quantum corrections \cite{deRham:2012ew} and strong coupling comes from studying the decoupling limit Lagrangian. The Lagrangian that focuses on the interactions of the helicity zero mode essentially encapsulates all of the nontrivial aspects of massive gravity that make it different from General Relativity. For instance the evasion of the vDVZ discontinuity through the Vainshtein mechanism is best understood in the decoupling limit theory. In the decoupling limit, the helicity-0 mode of the massive spin-2 field behaves like a Galileon.  Galileons were originally discovered in an extra dimensional context in the DGP braneworld model \cite{Dvali:2000hr} and encapsulate  the Vainshtein screening mechanism \cite{Nicolis:2004qq, Nicolis:2008in} (for reviews on their effects and roles in dRGT, see \cite{deRham:2012az,Babichev:2013usa}). It is this screening that allows dRGT to evade the vDVZ discontinuity.  But since the helicity-0 mode is screened by the matter through the Vainshtein mechanism, the theory is consistent with current constraints from solars system tests and other tests of classical gravity \cite{pulsars1,pulsars2}. The decoupling limit has been used to show that there may exist solutions of massive gravity which exhibit superluminal group velocities \cite{Nicolis:2008in,Adams:2006sv}, which extend into the full theory \cite{Deser:2012qx}, the instability of many of these solutions \cite{Berezhiani:2013dca}, and it has also been used to explain how superluminal group velocities can be consistent with causality \cite{Burrage:2011cr}. 

While the helicity-0 interactions in the decoupling limit are well-understood, the description of the interactions of the helicity-1 mode in the decoupling limit are not. We perform a \stu analysis of dRGT that allows us to develop a method for resumming the interactions of the helicity-1 and helicity-0 Goldstone-\stu fields to all orders in the decoupling limit (for work in the metric language see \cite{Koyama:2011wx}). For instance, this could be relevant for the study of a potential candidate for a Partially Massless theory of gravity, see Refs.~\cite{Deser:1983mm,Deser:2001pe,Deser:2001us,Deser:2001xr,Deser:2004ji,Zinoviev:2001dt,Zinoviev:2006im,Deser:2006zx,Joung:2012hz,deRham:2012kf,Deser:2013xb,Deser:2013uy,Hassan:2012gz,deRham:2013wv}. Furthermore, when extending this analysis on Anti-de Sitter the role of the vectors could be particularly interesting as previously pointed out in \cite{deRham:2012kf}.

We begin in section 2 with a review of the \stu procedure for massive gravity.  We exploit the vierbein formulation of the action including \stu fields for both diffeomorphisms and local Lorentz transformations. We clarify the role that the Deser-van Nieuwenhuizen (DvN) condition (also referred to as the ``symmetric vierbein condition'') plays in dRGT and show how this arises as the equation of motion for the Lorentz \stu field. We also clarify the relation with the metric formulation of massive gravity and give an explicit solution for the Lorentz \stu field that enforces the symmetric vierbein condition. In section 3, we directly perform a decoupling analysis on the Goldstone-\stu fields.  We determine the equations of motion and solution for the Lorentz \stu field in the decoupling limit. This solution allows us to write down the resummed interactions to all orders in the decoupling limit.  While this work was in preparation, the following article appeared which has some overlap in discussion \cite{Gabadadze:2013ria}.


\section{\stu Procedure for dRGT}
\label{sec: nonlinstuck}

In this section we will explore the \stu procedure in the second order Einstein-Cartan (vierbein) formulation of dRGT massive gravity about flat spacetime (see \cite{deRham:2012kf} for its decoupling limit on (Anti-)de Sitter). We shall focus our attention on the theory in four dimensions, the generalization to general dimensions being straightforward. Since four dimensional General Relativity in Einstein-Cartan form admits two separate local symmetry groups, 4 diffeomorphisms and 6 local Lorentz transformations, and since all 10 of these symmetries are broken in massive gravity, it is natural when analyzing the physics of massive gravity to reintroduce all 10 of these symmetries using the \stu formalism. 

In subsection 2.1, we will restore Lorentz invariance to the vierbein with a Lorentz \stu or compensator field. In subsection 2.2, we integrate out these auxiliary Lorentz \stu field.  In subsection 2.3, we will solve the equations of motion using Lorentz \stu formalism and substitute back into the action, thus re-arriving at the fully gauge-invariant dRGT action in metric language. Finally, we note the importance of the DvN gauge and the role it plays in the independence of the Lorentz \stu field's equations of motion from the $\b$ coefficients.

\subsection{\stu Fields for the Vierbein Formalism}

We begin with the action for massive gravity in the vierbein formalism:
\ba
S_{\rm vierbein} &=& \frac{\mpl^2}{2} \ep_{abcd} \int \left[ \frac{1}{4}E^a \wedge E^b \wedge R^{cd} \right.  \\
&& -m^2 \frac{\b_0}{4!} E^a \wedge E^b \wedge E^c \wedge E^d  \nonumber \\
&& -  m^2 \frac{\b_1}{3!} I^a \wedge E^b \wedge E^c \wedge E^d \nonumber \\
&& - m^2 \frac{\b_2}{2!2!} I^a \wedge I^b \wedge E^c \wedge E^d \nonumber \\
&& \left. - m^2 \frac{\b_3}{3!} I^a \wedge I^b \wedge I^c \wedge E^d \right]\,, \nonumber
\ea
where $I^a = \de^a_{\mu}dx^{\mu}$ and $ E^a = E^a_{\mu} dx^{\mu} $.
The conversion between the $\b$ and $\al$ (see equation (\ref{eq:metricaction}) that accounts for the tadpole cancellation conditions (which ensures a Minkowski vacuum) are:
\ba
 \b_0 &=& -12 - 8\al_3 - 2 \al_4  \\
 \b_1 &=& 6 + 6\al_3 + 2 \al_4 \nonumber \\
 \b_2 &=& -2 -4\al_3 - 2\al_4 \nonumber \\
 \b_3 &=& 2\al_3 + 2\al_4\,.\nonumber
\ea
Now we employ the \stu trick to this theory in order to restore gauge invariance under local Lorentz transformations (LLTs) and diffeomorphism (diff) transformations.  In order to accomplish this, we must introduce a Lorentz \stu field, $\Lambda^a{}_{b}$, and in addition \stu fields for diffeomorphisms in the combination $\frac{\p x^{\mu}}{\p \phi^{\nu}}$.  Thus we must make the following substitution into the action:
\be
E^a_{\mu} \to \La^a_{ \hspace{.2cm} b} E^b_{\nu} \frac{\p x^{\nu}}{\p \phi^{\mu}} \, .
\ee
With these extra fields, we gain back full LLTs and diff invariance, so we may freely move the diff \stu onto the background vierbein via compensating inverse diff:
\ba
E^a_{\mu} & \to & \La^a_{ \hspace{.2cm} b} E^b_{\nu} \nonumber \\ \de^a_{\mu} & \to & F^a_{\mu} = \frac{\p \phi^{\nu}}{\p x^{\mu}} \de^a_{\nu} = \p_{\mu}(\phi^a)\,.
\ea
Now the reference vierbein has gauge invariance under an independent set of diffs and LLTs, which renders the action gauge invariant.  In this framework we obtain new equations of motion from varying with respect to our new fields.  In the case of the diffeomorphism \stu fields, they are dynamical because they explicitly have to enter with derivatives.  As is well known, the helicity-0 mode of this field is the Galileon which comes with interactions with the helicity-2 sector of the massive spin-2 field, \cite{deRham:2010ik}.  They do not represent new degrees of freedom, of course, because they can be gauged away ($d\phi^a = I^a$) to return to the original action; they are therefore `borrowed' degrees of freedom from the massive spin-2 field. The virtue of the \stu formalism is it makes the different dynamics of the helicity-1 and helicity-0 modes manifest and easier to analyze than in the unitary gauge formalism in which the \stu fields are turned off (by definition unitary gauge is the gauge for which $\phi^a=x^a$ and $\Lambda^a{}_b=\delta^a{}_b$).
Contrarily, the \stu fields associated to local Lorentz invariance are auxiliary fields because the action does not depend on derivatives of them. Ignoring the kinetic terms and the cosmological constant $\b_0 $-term which are manifestly invariant under diffs and LLTs, we see that the only interactions with \stu fields present are:
\ba
S_{\rm mass} &=& - \frac{1}{2}\mpl^2 m^2 \e_{abcd}\int  \left[ \frac{\b_1}{3!}  F^a \wedge \( \La ^b_{ \hspace{.2cm} b'} E^{b'} \) \wedge \( \La ^c_{ \hspace{.2cm} c'} E^{c'} \) \wedge \( \La ^d_{ \hspace{.2cm} d'} E^{d'} \) \right.  \\
&+&\frac{\b_2}{2! 2!}  F^a \wedge F^b \wedge \( \La ^c_{ \hspace{.2cm} c'} E^{c'} \) \wedge \( \La ^d_{ \hspace{.2cm} d'} E^{d'} \) + \left. \frac{\b_3}{3!}  F^a \wedge F^b \wedge F^c \wedge \( \La ^d_{ \hspace{.2cm} d'} E^{d'} \) \right] \,.\nonumber
\ea


\subsection{Equations of Motion for the Lorentz \stu}
\label{sec:lorentz}

Let us set aside for the moment the diffeomorphism \stu fields and focus on integrating out the Lorentz auxiliary fields. To do this, it is worth mentioning a very powerful trick for describing the interactions of the original action, $S_{\rm metric}$, in terms of a deformed determinant \cite{Hassan:2011vm}.  A deformed determinant has all the same terms generated by expanding the determinant of a matrix, but each term is weighted by a different coefficient, $c_n$:
\ba
\mathcal{L}_{\rm mass} &= & - \frac{1}{2}\mpl^2 m^2 \widehat{\rm{Det}}[ \La E - F ]  \\
&=& c_0  \ep_{abcd} (\La E)^a \wedge (\La E)^b \wedge (\La E)^c \wedge (\La E)^d    \nonumber \\
& +& c_1 \ep_{abcd} F^a \wedge (\La E)^b \wedge (\La E)^c \wedge (\La E)^d \nonumber \\
& +& c_2 \ep_{abcd} F^a \wedge F^b  \wedge(\La E)^c \wedge (\La E)^d  \nonumber \\
& +& c_3 \ep_{abcd} F^a \wedge F^b \wedge F^c \wedge (\La E)^d \nonumber \\
& +& c_4 \ep_{abcd} F^a \wedge F^b \wedge F^c \wedge F^d \,,\nonumber
\ea
with $ (\La E)^a = \La^a {}_b E^b $ and comparing with the previous definition,
\be
c_n =  - \frac{1}{2 \, n! (4-n)!}\mpl^2 m^2 \beta_n \, .
\ee
Note that the term $c_4$ multiplies is just a shift in the action by a `constant', so we can consistently disregard it; likewise, $c_0$ being a cosmological constant term is manifestly invariant under LLTs so we may disregard it when integrating out $ \La^a {}_b$.

Using the Lorentz invariant properties of the Levi-Civita symbol and ${\rm Det}[\La]=1$, we may equivalently write this as
\ba
\mathcal{L}_{\rm mass} &= &  - \frac{1}{2}\mpl^2 m^2\widehat{\rm{Det}}[ \La E - F ]= - \frac{1}{2}\mpl^2 m^2\widehat{\rm{Det}}[  E -\La^{-1} F ]  \nonumber \\
&=& c_0  \ep_{abcd} E^a \wedge  E^b \wedge E^c \wedge  E^d     \\
& +& c_1 \ep_{abcd} (\La^{-1} F)^a \wedge E^b \wedge E^c \wedge  E^d \nonumber \\
& +& c_2 \ep_{abcd} (\La^{-1} F)^a \wedge (\La^{-1} F)^b  \wedge E^c \wedge E^d  \nonumber \\
& +& c_3 \ep_{abcd} (\La^{-1} F)^a \wedge (\La^{-1} F)^b \wedge (\La^{-1} F)^c \wedge  E^d \nonumber \\
& +& c_4 \ep_{abcd} (\La^{-1} F)^a \wedge (\La^{-1} F)^b \wedge (\La^{-1} F)^c \wedge (\La^{-1} F)^d \,. \nonumber
\ea
We may also express the deformed determinant as
\ba
\mathcal{L}_{\rm mass} &=& \sum_{n=0}^4 \frac{(-1)^n}{n!} c_n \frac{\partial^n}{\partial \mu^n} \,  {\rm{Det}}[ \La E - \mu \, F ] \big|_{\mu =0} \, .
\label{eq:det1}
\ea
Varying this Lagrangian with respect to the Lorentz auxiliary field, we find that the resulting equation for the \stu fields is independent of the $c_n$, i.e the $\beta_n$ constants. To see this, it is sufficient to show that if we take the Lagrangian to be ${\rm{Det}}[ \La E - \mu \, F ]$, the equation for $\La^a {}_b$ is independent of $\mu$ due to equation (\ref{eq:det1}).

Explicitly we find
\ba
\delta \, {\rm{Det}}[ \La E - \mu \, F ] &=& {\rm{Det}}[ \La E - \mu \, F ] \, {\rm Tr}\left[ \delta \Lambda E \, ( \La E - \mu \, F )^{-1} \right] \nonumber \\
&=& {\rm{Det}}[ \La E - \mu \, F ] \, {\rm Tr }\left[( \delta \Lambda \Lambda^{-1}\eta) \eta (\Lambda E) \, ( \La E - \mu \, F )^{-1} \right]\,,
\ea
where we have used matrix notation. It is understood that vierbein indices contract together and spacetime indices contract together and $\eta$ is the Minkowski metric which acts on vierbein indices. Using the property of Lorentz transformations $\Lambda \eta \Lambda^T = \eta$ we find
\be
\left( (\delta \Lambda) \Lambda^{-1} \eta \right)^T =- \left( (\delta \Lambda) \Lambda^{-1} \eta \right)\,.
\ee
This antisymmetry condition then imposes a symmetry condition on what it multiplies so that the trace vanishes
\be
{\rm Tr }\left[( \delta \Lambda \Lambda^{-1}\eta) \eta (\Lambda E) \, ( \La E - \mu \, F )^{-1} \right] =0 \, .
\ee
This gives the equation of motion for the Lorentz \stu fields 
\be
\eta (\Lambda E) \, ( \La E - \mu \, F )^{-1}  =\left[ \eta (\Lambda E) \, ( \La E - \mu \, F )^{-1} \right]^T = \left[ (\La E)^T - \mu \, F^T \right]^{-1} (\Lambda E)^T \eta \, ,
\ee
which is
\be
(\Lambda E)^T \eta ( \La E - \mu \, F ) = ( (\La E)^T - \mu \, F^T ) \eta (\Lambda E) \, .
\ee
It is easy to see that this reduces to the $\mu$ independent equation
\be
(\Lambda E)^T \eta F = F^T  \eta (\Lambda E)  \,,
\ee
or equivalently in tensor notation as
\be
(\Lambda E)^a_{\mu} \eta_{ab}F^b_{\nu} = (\Lambda E)^b_{\mu} \eta_{ab}F^a_{\nu} \, .
\ee
This is nothing more than the so-called DvN gauge condition between the two vierbeins $\Lambda E$ and $F$,   \cite{Hoek:1982za,Deffayet:2012zc}, sometimes also referred to as the ``symmetric vierbein condition'', \cite{Hinterbichler:2012cn} due to the fact that it symmetrizes over the spacetime indices.

Crucially this equation does not depended on $\mu$ and therefore it does not depend on the specific mass terms we have chosen! This may be verified by a brute force calculation for each individual term. Rather than a constraint, this equation is just the equation of motion for the non-dynamical Lorentz \stu fields. Solving this equation of motion explicitly in the decoupling limit will be the essential trick in obtaining the complete helicity-1 decoupling limit interactions. For now we content ourselves with the formal exact solution.


\subsection{Recovering the metric formulation}
\label{sec:lorentz2}

We now show that we can solve explicitly for the Lorentz \stu fields and on substituting back into the action show the complete equivalence with the metric form of dRGT massive gravity, crucially without needing to fix a gauge!\footnote{This approach follows the prescient work of \cite{Nibbelink:2006sz}.}

Given the relation
\be
(\Lambda E)^T \eta F = F^T  \eta (\Lambda E)\,,
\ee
this is equivalent to
\be
 \eta \Lambda^T \eta F E^{-1} = \eta (E^T)^{-1} F^T  \eta \Lambda\,.
\ee
Now using the property of Lorentz transformations $\Lambda \eta \Lambda^T  = \eta$ we can rewrite this as
\be
( \eta (E^T)^{-1} F^T  \eta \Lambda)^2 = \eta (E^T)^{-1} F^T  \eta \Lambda \eta \Lambda^T \eta F E^{-1}  =\eta (E^T)^{-1} F^T  \eta F E^{-1}\,.
\ee
This may be formally solved as\footnote{We should stress that the square root of a matrix is not unique and there are instances and backgrounds where it may be convenient to choose different soutions for the square root \cite{Deffayet:2012zc}, \cite{Gratia:2013gka}. For instance in square rooting a diagonal matrix we may choose to use different signs for the square root of each of the eigenvalues.
However, in most situations of physical interest, and certainly the regime covered by the decoupling limit,  $g^{-1} f$ has nonzero positive eigenvalues and the square root can be performed in the diagonalized basis with positive sign choice. } 
\be
\eta (E^T)^{-1} F^T  \eta \Lambda = \sqrt{\eta (E^T)^{-1} F^T  \eta F E^{-1} }\,.
\ee
Now performing a similarity transformation which acts as $E^{-1} \sqrt{Y} E = \sqrt{E^{-1} Y E}$,
\ba
E^{-1} (\eta (E^T)^{-1} F^T  \eta \Lambda) E &=&  E^{-1}  \sqrt{\eta (E^T)^{-1} F^T  \eta F E^{-1} } E \nn\\
&=& \sqrt{E^{-1}\eta (E^T)^{-1} F^T  \eta F  } \nn\\ 
&=& \sqrt{g^{-1}f}\,,
\ea
where we have defined the dynamical metric $g = E \eta E^T$ and reference metric $f = F \eta F^T$ in the usual way. Thus we see the appearance of the famous square root factor that is the building block of the dRGT action.

Finally the explicit formal solution for $\Lambda$ is
\be
\Lambda  = \eta (F^T)^{-1}g \sqrt{g^{-1}f} E^{-1} \, .
\ee
In component form this is the statement that
\be
\Lambda^a{}_b = \eta^{ac}F_c^{\mu} g_{\mu \alpha} (\sqrt{g^{-1}f})^{\alpha}{}_\nu E^{\nu}_b \, .
\ee
Although correct, this result is of limited use because it requires us to know the same square root structure that makes the metric formalism difficult to work with. However it does allow us to see the complete equivalence with the metric formalism. This is easy to see using the same deformed determinant method. We need only show that the combination ${\rm{Det}}[ \La E - \mu \, F ] $ is equivalent to the determinant that generates the dRGT mass terms.
This follows since
\ba
 {\rm{Det}}[ \La E - \mu \, F ] &=& {\rm{Det}}[\eta (F^T)^{-1}g \sqrt{g^{-1}f}  - \mu \, F ] ={\rm{Det}}[F]{\rm{Det}}[f^{-1}g \sqrt{g^{-1}f}  - \mu  ]  \nonumber \\
 &=& \sqrt{{\rm{Det}}[-g]} {\rm{Det}}[1  - \mu \sqrt{g^{-1} f}  ]\,,
\ea
where we have made extensive use of the rules ${\rm Det}[AB] = {\rm Det}[A]{\rm Det}[B]$ and ${\rm Det}[\sqrt{A}] = \sqrt{\rm Det[A]}$.

We now see that applying the same deformed determinant method we will generate all of the dRGT mass terms by expanding the determinant in powers of $\mu$. This confirms the known equivalence between these two formalisms without the need to specify any gauge choice in the intermediate steps.

\subsection{Origin of Symmetric Vierbein Condition }
\label{sec: allint}

To recapitulate the point of the previous sections, we see that the equations of motion for the Lorentz \stu field is nothing more than the DvN condition phrased in terms of the two vierbeins $(\Lambda E)$ and $F$ regardless of the gauge choice. We see that this condition is completely independent of the values of the ${\al}$'s and is thus a universal feature of massive gravity.

Let us now run the argument in reverse in the gauge fixed form. For reference, the equation of motion of $ \La $ can be put in the form:
\be
(E^{-1}\La^{-1} \eta F) = (E^{-1} \La^{-1} \eta F)^T\,. \nonumber
\ee
Up to now we have not made any specific gauge choice and we may choose one that seems convenient. The simplest gauge choice is to use a Lorentz transformation to set $\Lambda=1$. We then find the condition
\be
(E^{-1}\eta F) = (E^{-1}\eta F)^T\,, \nonumber
\ee
as a consequence of the equations of motion for the Lorentz \stu fields. If in addition we choose $F$ to correspond to Minkowski spacetime and choosing unitary gauge $\phi^a = x^a$, then $F=I$ and so this condition implies
\be
(\eta E)^T = (\eta E) \,,
\ee
which is the so-called symmetric vierbein condition $E_{\mu a}=E_{a \mu}$.

Alternatively we may impose as a {\it gauge choice} the DvN gauge written in terms of $E$ and $F$:
\be
(E^{-1}\eta F) = (E^{-1}\eta F)^T,
\ee
which then implies that as a consequence of the equations of motion for the Lorentz \stu fields $\Lambda = 1$. In this representation we need to worry about whether it is possible to choose this gauge. In the previous representation it is clear that we can always use a Lorentz transformation to set a Lorentz \stu to unity.

Following the previous reasoning, working in the gauge $\Lambda = 1$ for general $F$ since $ g^{-1} = E^{-T}\eta E^{-1} $ and $ f = F^T \eta F $, together these imply:
\be
\sqrt{g^{-1}f} = \sqrt{E^{-T}\eta E^{-1} F^T \eta F} = \sqrt{E^{-1}\eta F E^{-1} \eta F} = E^{-1}\eta  F  \, .
\ee
Substituting this value in dRGT in the metric language, we see that the metric dRGT action is equivalent to the vierbein action\footnote{Here we are using the $\b$ coefficients and $ e_i \( \sqrt{g^{1}f} \) $ for the action, which is equivalent to $\al$'s with $ e_i( \mathcal{K} ) $ after the tadpole cancellation conditions are applied.} and has the form:
\ba
S_{\rm metric} &\equiv& - \frac{\mpl^2 m^2}{2}\int  \sqrt{-g}\, \sum  \b_i \, e_i \( \sqrt{g^{-1}f  } \)\nn \\
 &=& -\frac{\mpl^2 m^2}{2} \int {\rm det} E\,  \sum \b_i \, e_i( E^{-1}F ) \equiv S_{\rm vierbein} \,.
\ea
This explicitly shows the connection between the vierbein and metric formulations of dRGT in the gauge fixed form.

We see that the \stu formulation significantly clarifies the role of this condition in the vierbein formalism and the connection with the metric formulation. The reason why we recover the DvN condition by integrating out the Lorentz \stu fields is that it is this gauge-fixing that put it into that form in the first place. Since the gauge-fixing is independent of the mass parameters, so too must the equation of motion that restores the gauge symmetry.

Returning to the issue of the decoupling limit, one can check that trying to perturbatively solve in powers of $m$ and $\mpl$ for the helicity-1 interactions from the square root structure is somewhat unwieldly since there is no simple expression for the square root of a matrix without going through the process of diagonalization (see \cite{Koyama:2011wx} for an alternative approach which is amenable to perturbative theory in the metric language).  However, exploiting the vierbein's ability to make a simple form for the metric square root structure, we are able to find a meaningful resummation of the interactions.  

\section{Decoupling Limit Interactions }
\label{sec: allint}

\subsection{Scaling}

We now move to the calculation of the full form of the interactions in the decoupling limit. To compute the interactions for helicity-1 field to all orders, we will have to integrate out the Lorentz auxiliary field, and substitute the solution back into the action. We have already done this formally in section \ref{sec:lorentz2} but this expression was difficult to work with. We will now see how to solve this in the decoupling limit and generate the infinite number of terms in the action, which have to be formally resummed. We will use the following canonically normalized expressions for the helicity-2 mode and St\"uckelberg fields
\ba
E^a_{\mu} &=& \de^a_{\mu} + \frac{1}{2 \mpl } e^a_{\mu} \nonumber \\
\Lambda^a{}_{b} &=& e^{ \hat{\om}^{a}{}_{ b}} = I + \hat{\om}^{a}{}_{ b} + \frac{1}{2} \hat{\om}^{a}{}_{c} \hat{\om}^{c}{}_{b} + \cdots  \\
\p_{\mu}\phi^a &=& \p_{\mu} \( x^a + \frac{B^a}{m \mpl} + \frac{E^{a \nu} \p                                                      _{\nu}\pi}{ \Lambda^3_3} \)\nonumber
\ea
and perform the scaling or decoupling limit,
\ba
\mpl   \to  \infty \,, \hspace{20pt} m \to  0 \hspace{20pt}\text{while keeping}\hspace{20pt}  \Lambda_3 = (m^2\mpl)^{\frac{1}{3}} \to \rm{const}\,.
\ea 

In addition $\hat{\om}^{a} {}_b$ should have some scaling with respect to $m$ and $\mpl$.  By inspecting the form of the action, or the equation of motion for the Lorentz \stu fields,  we will see that the only way it can survive the limit successfully (i.e. generating no divergent terms) is with the following normalization:
\be
\hat{\om}^{a}_{\hspace{.2cm} b} = \frac{\om^{a}_{\hspace{.2cm} b}}{m \mpl}\,.
\ee
This scaling is easy to understand, the antisymmetry of the $\hat{\om}^{a}{}_b$ implies that the leading order contribution from $\pi$ vanishes and that $\hat{\om}^{a}{}_b$ is principally determined by the vector fields $B^a$ which come normalized with a factor of $1/(m \mpl)$.

Since the equation of motion for $\om$ must be independent of ${\alpha}$'s, we can select whichever form of the dRGT action we like to determine $\om$. Alternatively we may directly use the equation of motion for the $\Lambda$'s derived in section \ref{sec:lorentz}. To explain the steps used in deriving the decoupling limit, for now we show how to determine the \stu fields directly from the action. For simplicity, we will use the St\"uckelberg'ed $\beta_1$ term of $S_{\rm mass}$, namely the minimal model (up to a cosmological constant term):
\be
\mathcal {L} =-\beta_1 \frac{m^2 \mpl^2}{12} \left[ \ep_{abcd} e^{ \hat{ \omega}^a {}_{a'}  } E^{a'} \wedge e^{ \hat{\omega}^b {}_{b'}  } E^{b'} \wedge e^{  \hat{\omega}^c{} _{c'} }E^{c'} \wedge d \( x^d + \frac{B^d}{m\mpl} + \frac{\p^d \pi}{\Lambda^3_3}\) \right] \nonumber
\ee
\ba &=&-\beta_1 \frac{m^2 \mpl^2}{12}  \ep_{abcd}\( \de^a_{a'} + \hat{\om}^{a} {}_{a'} + \frac{1}{2}  \hat{\om}^{a}{}_m \hat{\om}^m {}_{a'} + \cdots \) E^{a'} \wedge  \( \de^b _{b'} + \hat{\om}^{b} {}_{b'} + \frac{1}{2} \hat{\om}^{b} {}_m \hat{\om}^m {}_{b'} + \cdots \) E^{b'}  \nonumber \\
&& \wedge  \( \de^c_{c'} + \hat{\om}^{c} {}_{c'} + \frac{1}{2} \hat{\om}^{c} {}_m \hat{\om}^m {}_{c'} + \cdots \) E^{c'} \wedge \( I^d + \frac{\d B^d}{m \mpl} + \frac{\d \( \p^d\pi \) }{\Lambda^3_3} \)\,.
\ea
To convert back to tensor notation, we may replace the wedge products with a Levi-Civita symbol contracting all spacetime and vierbein indices.  However, the two Levi-Civita symbols (one being for the Lorentz indices, the other being for the spacetime indices) can be combined to create a generalized Kronecker delta symbol\footnote{To be clear on conventions, we take the standard definitions of the Kronecker deltas: $ \de^{\mu \nu \rho \sigma}_{abcd} = \ep^{\mu \nu \rho \sigma}\ep_{abcd} $.  This means that the Kronecker deltas are not normalized to be `weight one'. More generally we have $\de^{\mu \nu \rho}_{abc} = \frac{1}{1!} \ep^{\mu \nu \rho d}\ep_{abcd}$ and $\de^{\mu \nu}_{ab} = \frac{1}{2!} \ep^{\mu \nu c d}\ep_{abcd}$}.  In all of what follows, the generalized Kronecker delta will be symbolized by an $\ep$ without indices and a $\de^{\mu\nu\rho\si}_{abcd}$ with explicit indices. The action then takes on the form
\be
= -\beta_1 \frac{m^2\mpl^2}{12}  \ep \(I+ \frac{\p B}{m \mpl} + \Pi \) \(I + \frac{e}{2 \mpl} + \frac{\om}{m \mpl} + \frac{\om \cdot \om}{2m^2\mpl^2} + {\mathcal O}(\mpl^{-3/2}) \)^3\,.
\ee
Here the indices have been suppressed, because objects commute freely under generalized Kronecker delta, which now holds all of the index structure, and we have defined
\ba
&& \om \cdot \om = \om^{a}_{\hspace{.2cm} c} \om^{c}_{\hspace{.2cm} b} \\
&& \Pi = \frac{\p_{\mu} \p^a \pi}{\La^3_3}\,.
\ea
This allows us to expand and collect the relevant terms of the Lagrangian more easily. Note that we do not need to be concerned with the $\Pi$'s dimensions because it survives the decoupling limit, so we have just absorbed it into the definition of $\Pi$.

Now we must expand out the Lagrangian. There will be many terms, but since we are only interested in the ones that will contribute to the equations of motion of the Lorentz \stu field, we may ignore the helicity-0 kinetic interactions, the helicity-0-helicity-2 interactions, the total derivatives, terms that manifestly disappear in the decoupling limit (i.e. $O(\frac{1}{\mpl^3})$ and higher), and the terms that are annihilated by the tadpole cancellation condition.  In fact, relative to the overall $ m^2\mpl^2 $  its easy to verify that:

\begin{enumerate}

\item[1.)] The highest order of $\om$ that matters are $\om^2$ terms, all higher order can be consistently neglected because they contain too many powers of inverse $\mpl$.

\item[2.)] We can consistently disregard the helicity-2 mode, its $\frac{1}{\mpl}$ scaling never allows to have an interaction with $\om$ that survives the decoupling limit. It interacts only directly with $\pi$ through known structures.

\item[3.)] Any terms of the form $\( \frac{1}{m \mpl}\)^n$ with $n>1$ also cannot couple because they will vanish in the decoupling limit.

\end{enumerate}
This leaves only 2 kinds of terms.  Firstly, the terms that do not scale with $\mpl$, which must couple to terms like $\frac{\om^2}{m^2 \mpl^2 }$; secondly, the terms that scale as $\frac{1}{m \mpl}$, which must couple to terms like $\frac{\om}{m \mpl}$.  We can easily read off that these combinations --and only these two combinations-- cancel off the overall $ m^2\mpl^2 $ prefactor.

\subsection{Determining Lorentz \stu from action}

Using the logic of section 3.1, we find that we can consistently truncate the relevant portion of the action to the following:
\ba
&&-\beta_1 \frac{m^2\mpl^2}{12}  \ep \(I + \frac{\p B}{m \mpl} + \Pi \) \(I^3 + 3 \( \frac{I^2 \om}{m \mpl} + \frac{I \om^2}{m^2 M^2_{pl}} \) + 3 \frac{ I^2 \om \cdot \om}{2m^2M^2_{pl}} \) \nn \\
&=& -\beta_1 \frac{1}{4}  \ep \( \p B I^2 \om + (I + \Pi) \( I \om^2 + \frac{I^2 \om \cdot \om}{2} \)  \) + {\mathcal O} \( \frac{1}{\mpl^{1/2}} \) \\
&=& -\beta_1 \frac{1}{4} \de^{\mu \nu \rho \sigma}_{abcd} \( \p_{\mu} B^a \de^b_{\nu} \de^c_{\rho} \om^d_{\hspace{.2cm} \si} + (\de + \Pi)^a_{\mu} \( \de^b_{\nu} \om^c_{\rho} \om^d_{\si} + \frac{\de^b_{\nu} \de^c_{\rho} \om^{d}_{\hspace{.2cm} c} \om^{c}_{\hspace{.2cm} \si} }{2} \)  \)\,.\nn 
\ea
Where in the last line we restored all of the indices\footnote{Since we are in the Minkowski space decoupling limit, we will make no distinction between Lorentz indices $\{ a,b,c, \dots \}$ and spacetime indices $ \{ \mu,\nu,\rho,\dots \}. $}. In obtaining this result we have discarded all terms which are total derivatives. Now we can vary the action and obtain the equation of motion for $\om^{a}_{\hspace{.2cm} b}$.  Firstly, we must note that $\om \in \frak{so}(1,3)$ and therefore $\om_{ab} = -\om_{ba}$.  This affects our equations of motion through the variational parameter\footnote{Our convention is always `weight-one' anti-symmetrization.  Therefore, $A_{[ab]} = \frac{1}{2} \( A_{ab} - A_{ba} \).$}:
\be
\frac{\de \mathcal{S}}{\de \om^{ab} }\de \om^{ab} = A_{ab}\de \om^{ab} = A_{ab}\de \om^{[ab]} = A_{[ab]}\de \om^{ab} \\ \Rightarrow A_{[ab]} = 0\,,
\ee
or, if one prefers, in terms of a functional derivative,
\be
\frac{\de \om_{ab}}{\de \om_{cd}} = \frac{1}{2} \de^{cd}_{ab}.
\ee
Factoring out a common $\de$, the equation of motion now schematically looks like:
\be
\ep\ \Big( \underbrace{G \de \frac{\de \om}{\de \om} \eta}_{(A)} + (\de + \Pi) \Big( \underbrace{2 \om \frac{\de \om}{\de \om}}_{(B)} + \underbrace{\de \frac{\de \om}{\de \om} \cdot \om}_{(C)} + \underbrace{\om \cdot \frac{\de \om}{\de \om}\de}_{(D)}  \Big) \Big) = 0\,.
\ee
We will split up the calculation and the Kronecker deltas into their following parts:
\ba
(A) &=& \de^{\mu\nu\rho}_{abc} \de^a_{\rho} \p_{\nu}B^b \de^{a\mu'}_{\al \b} \eta_{\mu' \mu} \nonumber \\ &=& 2 G_{\al \b} \\ (B) &=& 2 \de^{\mu\nu\rho}_{abc} \om^a_{ \hspace{.1cm} \mu} \de^{b \nu}_{\al \b} \eta_{\nu\nu'} (\de + \Pi)^c_{\rho} \nonumber \\ &=& 4 \left[ (2 + [\Pi])\om_{\al \b} + \om_{a \al} \Pi^{a}_{\b} - \om_{a\b} \Pi^{a}_{\al} \right] \\ (C) &=& \de^{\mu\nu}_{ab} \de^{a c}_{\al \b} \om_{c\mu} (\de + \Pi)^b_{\nu} \nonumber \\ &=& -2 \om_{\al \b} \(3 + [\Pi] \) - \( \om_{\b a} \Pi^{a}_{\al} - \om_{\al \b} \Pi^{a}_{\b} \) \\ (D) &=& \de^{\mu\nu}_{ab}  \om^a_{ \hspace{.1cm} \gamma} \de^{\gamma\mu'}_{\al \b} \eta_{\mu\mu'} (\de + \Pi)^b_{\nu} \nonumber \\ &=& (C)\,,
\ea
where $[\Pi] = \Pi^a_a$ and  $G_{\al \b} = \p_{\al} B_{\b} - \p_{\b} B_{\al}$.  Finally, when combined they yield the following equation of motion:
\ba
-(A) &=& (B) + (C) + (D) \nonumber \\ \Longrightarrow G_{ab} &=& 2 \om_{ab} - ( \om_{ca} \Pi^c_b - \om_{cb} \Pi^c_a ) \, .
\ea

\subsection{Determining Lorentz \stu from equation of motion}

The simplicity of this result suggests that there is a quicker way to derive it, and that is indeed the case. We can use the formal equation of motion derived in section \ref{sec:lorentz} remembering that $F= \partial \phi$, i.e. $F^a_{\mu} = \partial_{\mu} \phi^a$,
\be
(\Lambda E)^T \eta \partial \phi = (\partial \phi)^T  \eta (\Lambda E)
\ee
which amounts to
\be
E^{\mu}_a \left( e^{\hat \om} \right)_{bc} \partial_{\mu}\phi^c = E^{\mu}_b \left( e^{\hat \om} \right)_{ac} \partial_{\mu}\phi^c \, .
\ee
We now expand out this expression to leading nonzero order in the decoupling limit which turns out to be order $1/(m \mpl)$ to give
\be
\om_{b}{}^c \left( \delta_{ca} + \Pi_{ca} \right) + \partial_{a} B_b = \om_{a}{}^c \left( \delta_{cb} + \Pi_{cb} \right) + \partial_{b} B_a  + \dots
\ee
or equivalently,
\be
\om_{ba} + \om_{bc} \Pi^c_{a} + \partial_{a} B_b =   \om_{ab} + \om_{ac} \Pi^c_{b}  + \partial_{b} B_a \, .
\ee
Rearranging and using the various symmetry properties this gives
\be
G_{ab} = 2 \omega_{ab} - ( \om_{ca} \Pi^c_b - \om_{cb} \Pi^c_a )\,.
\ee
which is the desired relation.

\subsection{Nonlinear parameterization}

Whilst there is not a simple expression for the solution of this equation, we can at least always invert these equations for $\om_{ab}$ since we can solve the 6 linear equations for the 6 unknowns.  Fortunately however, there is a way to write a formal solution to this equation that does not require inverting a $6 \times 6$ matrix.  To see this, we will utilize the following trick: at a given point in spacetime, we can always use a global Lorentz transformation to diagonalize $\Pi_{ab}$ at that point so that
\be
\Pi_{ab} = 0  \text{ if }\ a \neq b\,.
\ee
This is always possible because a symmetric tensor can always be diagonalized with a Lorentz transformation.  Then we see that the equation is can now be inverted at that point:
\ba
 G_{ab} = \( 2 + \Pi^a_a + \Pi^b_b \) \om_{ab} \\
\Longrightarrow \om_{ab} = \( \frac{G_{ab}}{2 + \Pi^a_a + \Pi^b_b} \)\,.
\ea
Here $\Pi^a_a$ denotes the diagonal components of $\Pi$, and we do not sum over $a$. 

Now, we would like to covariantize this expression to give one that is valid at all points in spacetime even when $\Pi_{ab} $ is not diagonal. The covariantization can be accomplished by employing a Schwinger-parameterization technique:
\be
\om_{ab} =\int^{\infty}_{0} \d u \hspace{.2cm} e^{-2u} e^{ -u\Pi_a{}^{a'} } G_{a'b'} e^{ -u\Pi^{b'}_{\hspace{.2cm}b}} .
\ee
One can see that when the previous fixing is made, $\Pi_{ab} = 0 $ if $a \neq b$, this expression trivially reverts to the correct solution.  However, in this form, the equation is manifestly covariant, so it is valid everywhere and is therefore the general covariantized solution for $\om_{ab}$.

This formal expression can also be used as the starting point for the development of perturbations around a given background. For instance, if we choose to expand around a background with $\Pi=0$ then we can expand the exponentials and perform the integrals on each term
\be
\om = \sum_{n,m} \, \int^{\infty}_{0} \d u \hspace{.2cm} e^{-2u} \frac{1}{n!m!} (-1)^{n+m}u^{n+m} \, \Pi^n G \Pi^m \, .
\ee
The integral may then be explicitly formed to give
\be
\om =  \sum_{n,m} \,  \frac{(n+m)!}{2^{1+n+m}n!m!} (-1)^{n+m}\, \Pi^n G \Pi^m \, .
\ee
Whilst cumbersome, this is the best form we can give the answer in, unless it happens that $\Pi$ commutes with $G$ which in general is not the case.

When considering perturbations around a general background we can again use this same method. For the background we must find the exact solution,
\be
\bar G_{ab} = 2 \bar \omega_{ab} - ( \bar \om_{ca} \bar \Pi^c_b - \bar \om_{cb} \bar \Pi^c_a ) \,,
\ee
where $\bar{X}$ denotes background quantities. Then the equation for the linearized fluctuations is
\be
\delta G_{ab} = 2 \delta \omega_{ab} - ( \delta \om_{ca} \bar \Pi^c_b - \delta \om_{cb} \bar \Pi^c_a ) -( \bar \om_{ca} \delta \Pi^c_b - \bar \om_{cb} \delta \Pi^c_a )  \, .
\ee
This equation may then be formally solved using the same Schwinger parameterization trick (in condensed notation),
\be
\delta \om = \int^{\infty}_{0} \d u \hspace{.2cm} e^{-2u} e^{ -u\bar \Pi_a{}^{a'} } (\delta G - \bar \om \delta \Pi + \delta \Pi \bar \om)_{a'b'} e^{ -u\bar \Pi^{b'}_{\hspace{.2cm}b}}   \,,
\ee
and then expanded out giving the expression
\be
\delta \om =  \sum_{n,m} \,  \frac{(n+m)!}{2^{1+n+m}n!m!} (-1)^{n+m}\, \bar{\Pi}^n (\delta G - \bar \om \delta \Pi + \delta \Pi \bar \om) \bar{\Pi}^m \, .
\ee
The usefulness of these expressions depends on the context.

\subsection{Full Decoupling Limit Action}

To determine the final form of the action it is helpful to make again use of the deformed determinant method. With this trick we can express the mass terms by expanding an expression of the form ${\rm Det} [\Lambda E - \mu F] = {\rm Det} [E - \mu \Lambda^{-1} \partial \phi]$. It is straightforward to see that the contribution of the helicity-1 mode comes entirely from the term where $E=1$ and $(\Lambda^{-1} \partial \phi)$ is expanded to second order in $1/(m \mpl)$. In other words helicity-1 Lagrangian is determined entirely by expanding
\be
- \frac{1}{2}\mpl^2 m^2  \, {\rm Det} [1- \mu  (\Lambda^{-1} \partial \phi)]\,.
\ee
To the relevant order we have
\be
 (\Lambda^{-1} \partial \phi )=(1+ \Pi)+ \frac{1}{m \mpl} \left( -\om (1+ \Pi) + \partial B \right) + \frac{1}{2m^ 2\mpl^2} \omega^2 (1+ \Pi)- \frac{1}{m^ 2\mpl^2} \omega \partial B \dots
\ee
We can then express the determinant as 
\ba
&& {\rm Det} [1- \mu  (\Lambda^{-1} \partial \phi)] = {\rm Det} [Z_{\mu}] \times\\
 &&  {\rm Det} [1-Z_{\mu}^{-1} \mu  ( \frac{1}{m \mpl} \left( -\om (1+ \Pi) + \partial B \right) + Z_{\mu}^{-1} \mu \frac{1}{2m^ 2\mpl^2} \omega^2 (1+ \Pi)- Z_{\mu}^{-1} \mu \frac{1}{m^ 2\mpl^2} \omega \partial B]  \nonumber 
\ea
where $Z_{\mu} = 1- \mu  (1+\Pi)$. 
Next we expand out the determinant and use the fact ${\rm Det} [Z_{\mu}]$ is a total derivative and that the piece linear in $1/{m \mpl}$ will vanish because the $\partial B$ term is always a total derivative and the $\omega$ term vanishes by antisymmetry. Then we have for the mass term
\ba
{\mathcal L}_{\rm mass}&=&\sum_{n=0}^4 \frac{(-1)^n}{n!} \frac{1}{8\, n! (4-n)!} \beta_n  \frac{\partial^n}{\partial \mu^n} \left[ 
 \mu \, {\rm Det} [Z_{\mu}]  {\rm Tr}[Z_{\mu}^{-1}  \omega^2 (1+ \Pi)- Z_{\mu}^{-1} \frac{1}{2} \omega \partial B] \nonumber  \right. \\
 &-&  \mu^2 {\rm Det} [Z_{\mu}] {\rm Tr}[\left(Z_{\mu}^{-1}  \left( -\om (1+ \Pi) + \partial B \right)\right)^2] \nonumber \\
 &+&\left.  \mu^2 {\rm Det} [Z_{\mu}] \left( {\rm Tr}[(Z_{\mu}^{-1}  ( \frac{1}{m \mpl} \left( -\om (1+ \Pi) + \partial B \right))] \right)^2 \right] \Big|_{\mu =0} \, .
\ea
The derivatives are straightforward to perform. The terms of the form $(\partial B)^2$ all vanish upon integration by parts. The terms linear in $\partial B$ are all multiplied by anti-symmetric quantities which allows us to replace $\partial_a B_b \rightarrow \frac{1}{2} G_{ab}$. Finally, including the usual helicity-2/helicity-1 interactions we have neglected up to now, the full form of the action to all orders in the decoupling limit is:
\ba
\mathcal{S}_{\text{D.L.}} &=& \int \d^4x \, \frac{1}{8} h^{\mu \nu} \hat{\mathcal{L}}^{\al \b}_{\mu\nu} h_{\al\b} + \frac{1}{2}h^{\mu\nu} \( X^{(1)}_{\mu\nu} + \frac{1+ \al_3}{\La^3_3} X^{(2)}_{\mu\nu} + \frac{\al_3 + \al_4}{\La^6_3} X^{(3)}_{\mu\nu} \)  \\
&-&  \frac{\beta_1}{4} \de^{\mu\nu\rho\si}_{abcd}   \( \frac{1}{2} G_{\mu}^a \om^b{}_{\nu} \de^c_{\rho} \de^d_{\si} + (\de + \Pi)^a_{\mu}
 [ \de^b_{\nu} \om^c{}_{\rho} \om^d{}_{\si}
  +\frac{1}{2}\de^b_{\nu} {\de}^c_{\rho}  \om^d{}_{\al} \om^{\al}{}_{\si} ]
 \) \nonumber \\
&-& \frac{\beta_2}{8} \de^{\mu\nu\rho\si}_{abcd}  \left(  2 G_{\mu}^a (\de + \Pi)^b_{\nu} \om^c {}_{\rho}  \de^d_{\si}+  (\de + \Pi)^a_{\mu} (\de + \Pi)^b_{\nu} [  \om^c{}_{\rho}\om^d{}_{\si}+ \de^d_{\si}\om^c{}_{\al} \om^{\al}{}_{\rho}]   \right)\nonumber \\
&-&  \frac{\beta_3}{24} \de^{\mu\nu\rho\si}_{abcd}  \left( (\de + \Pi)^a_{\mu} (\de + \Pi )^b_{\nu} (\de + \Pi)^c_{\rho} \om^d{}_{\al} \om^{\al}{}_{\si} +3 \om^a {}_{\mu} G_{\nu}^b (\de + \Pi)^c_{\rho} (\de + \Pi )^d_{\si}  \right) \nonumber \, ,
\ea
where $\hat{\mathcal{L}}^{\al \b}_{\mu\nu}$ is the Lichnerowicz operator defined with the convention $\hat{\mathcal{L}} = \Box + \dots$ and we remind the reader of the known relations
\ba
\om_{ab}  &=&\int^{\infty}_{0} \d u \hspace{.2cm} e^{-2u} e^{ -u\Pi_a{}^{a'} } G_{a'b'} e^{ -u\Pi^{b'}_{\hspace{.2cm}b}} \\
&=&\sum_{n,m} \,  \frac{(n+m)!}{2^{1+n+m}n!m!} (-1)^{n+m}\, \( \Pi^n G \Pi^m \, \)_{ab}\,, \nonumber 
\ea
where once again $G_{ab}$ and $\Pi_{ab}$ are defined as
\ba
G_{ab}  = \partial_a B_b - \partial_b B_a \, , \hspace{20pt}{\rm and}\hspace{20pt} \, \Pi_{ab} = \frac{\p_a \p_b \pi}{\La^3_3}\,.
\ea
The transverse tensors $X^{(n)}\mn$ are given by \cite{deRham:2010ik},
\ba
{X^{(1)}}^a_{\mu} &=& - \delta^{ab}_{\mu \nu} \Pi^{\nu}_b  \nonumber \\
{X^{(2)}}^a_{\mu} &=& \frac{1}{2!} \delta^{abc}_{\mu \nu \rho} \Pi^{\nu}_b \Pi^{\rho}_c  \\
{X^{(3)}}^a_{\mu}  &=&  -\frac{1}{3!} \delta^{abcd}_{\mu \nu \rho \si} \Pi^{\nu}_b \Pi^{\rho}_c \Pi^{\si}_d  \, ,
\ea
and the $\beta$ and $\alpha$ coefficients are related by
\ba
\b_1 &=& 6 + 6\al_3 + 2 \al_4 \nonumber \\
\b_2 &=& -2 -4\al_3 - 2\al_4 \nonumber \\
\b_3 &=& 2\al_3 + 2\al_4 \, .\nonumber
\ea
Inspection of the above action shows it is manifestly $U(1)$ gauge invariant and that it is automatically quadratic in the gauge field $G_{ab}$, considering the vacuum solution $\Pi=0$ we see that we recover an ordinary Maxwell action for $B^a$.  Thus there are no $\text{vector-vector-vector}$ interactions, but there are an infinite number of ${\text{vector-vector-(scalar)}}^n$ interactions as already noted in \cite{Koyama:2011wx}.

\section{ Conclusions }
\label{sec: conclu}

We have given the complete expression for the decoupling limit of dRGT massive gravity, written down explicitly with all of the interactions amongst helicity-0, helicity-1 and helicity-2 fields.  We show how the resummation of the helicity-1 interactions is greatly simplified in the vierbein formalism with the inclusion of both diff and LLT \stu fields. We give an exact solution for the Lorentz \stu fields in the full nonlinear theory, which demonstrates the equivalence of the vierbein and metric language of dRGT. We have shown that the non-dynamical equations for the Lorentz \stu fields imposes the ``symmetric vierbein'' condition. It would be interesting to further explore the physical consequence of these helicity-1/helicity-0 interactions and also to explore the precise form of these interactions in theories with multiple gravitons/vierbeins. We leave these considerations to future work.

\acknowledgments

We thank Claudia de Rham, Matteo Fasiello, Gregory Gabadadze, Kurt Hinterbichler, David Pirtzkhalava and Andrew Matas for useful discussions. AJT is supported by DOE grant DE-FG02-12ER41810.




%
\end{document}